# TOWARDS AUTOMATED WEB APPLICATION LOGIC RECONSTRUCTION FOR APPLICATION LEVEL SECURITY

*George Noseevich and Dennis Gamayunov*
Information Systems Security Lab,
Lomonosov Moscow State University, Moscow, Russia

**Abstract**
Modern overlay security mechanisms like Web Application Firewalls (WAF) suffer from inability to recognize custom high-level application logic and data objects, which results in low accuracy, high false positives rates, and overhelming manual effort for fine tuning. In this paper we propose an approach to web application modeling for security purposes that could help next-generation WAFs to adapt to specific web applications, and do it automatically whenever possible. We aim at creating multi-layer models that adequately simulate various aspects of web application functionality that are significant for intrusion detection and prevention, including request parsing and routing, reconstruction of actions and data objects, and action interdependencies.

## 1. Introduction

This paper addresses the problem of detecting and preventing intrusions on web applications with the help of web application firewalls (WAF). This kind of dedicated security products evolved from generic intrusion detection systems (IDS) as the significance of web applications increase, the complexity level of the web applications grows, and the inadequacy of traditional IDS in detecting and protecting against intrusions on web applications becomes more recognized.

Despite there is a plenty of WAFs nowdays (both commercial and open source), empirical reasearch ([1], [2]) has shown that such firewalls do not provide satisfactory protection from modern intrusions, whereby today's commericial firewalls are currently losing out to open source firewalls ([4], [5]) and in some cases companies find themselves compelled to develop their own solutions for protecting web-applications from scratch ([6]). In analyzing today's web application firewalls, the following flaws should be noted that reduce their efficiency:

- today's versions of commercial WAF's from industry leaders are subject to variations of "old" traversing methods [1] on the HTTP message parsing level;

- in any substandard cases, in order for the an application to function normally, a significant portion of the protection mechanisms will have to be switched off. Reaching a configuration that provides a reasonable level of false responses along with protection from at least base intrusions (XSS, SQLi) is an arduous process that requires a months-long operation and hundreds of hours of man power;

- there is no protection against a broad range of intrusions on application logic (including intrusions on authorization mechanisms).

The main reason why today's web application firewalls are unable to effectively fight intrusions on web applications is that these firewalls perform intrusion detection mostly at HTTP protocol level. At the same time the web applications development technology has long since escaped the bounds of a "clean" HTTP: they construct multi-level abstractions based on it, and provide application authors with the tools to resolve typical tasks, such as authentication and session support, users authorization, structured data messaging, etc. Developing right behind the development of web technologies, firewalls have taken ad-hoc routes, adding support for separate new technologies (for example, REST applications) based on old logic or integrating new bundles of signatures in response to the publication of new intrusions or traversing methods.

As a result, the way of how WAF firewalls understand internals of applications and their functioning has proven to be far from realistic. In order to resolve this issue, firewalls should adapt to each of the applications they protect: the level of abstraction must be raised, along with complexity growth in web application development. The task is even more complex since there is no standard protocol for structured data exchange between the server side of the web application and the client side executed in the browser. Thus, each application implements its own protocol on top of the HTTP that differs from the others both syntaxically and semantically.

## 2. The Problem of Web Applications Modeling

In order to ensure the adequate protection of today's web applications, firewalls should primarily raise the level of "understanding" of how the protected objects are organized. From a formal standpoint, this requires contruction of *functional models*, which adequately and properly describe all properties of a web application relevant to the task of detecting intrusions. In this section we describe the requirements for such models – which of the web applications functioning aspects (properties) should be modelled.

### Conformity (Fitness)

Expressive power must be sufficient to describe a broad class of web applications, preserving a balance between the universality of the model and the rest of its properties. In general, web applications are arbitrary programs and an attempt to take into account all possible options of their implementation inevitably leads to models that are difficult to comprehend, computationally complex, and do not allow automation.

The sufficiency of the expressive power of a model may not be established in a formal fashion. As an alternative, sufficient conformity of the model may be declared with the help of:
- expert evaluation by specialists for the development of web applications as well as by specialists in web application protection analysis;
- creating models for building web applications for widespread frameworks.

### Automation possibility

Our final goal is automation of the web application model creation by the overlay security application. Otherwise, as we can draw from the current experience, expenses for the manual creation and maintenance of such models outweigh the benefit of the additional protection that they provide by several times. Furthermore, in cases where building the models does require the involvement of an operator, the required level of expertise should be minimized.

### Interpretability

Despite the need for models creation to be automated, the burden of manual analysis and correction in selecting the mathematical formalisms used in the models should be evaluated. Models which follow the "black box" principle such as neural networks will have a limited practical use since the average security analyst can not make any conclusions out of typical neural network recognition results – why did it classify the object one way or another, which specific data parts and values caused the anomaly, and so on.

## 3. Web application functional model synthesis

As we mentioned above, the creation of an efficient web application firewall requires the development of a web app model that takes all relevant aspects of its functioning into account. In this paper, we divided the modelling aspects into three levels from the lowest to the highest abstraction, and only the most important aspects will be described in detail. The names of the aspects are written in **bold**.

*3.1 Syntax level*

The syntax level is used to model the process of selecting a specific web application from a multitude of web applications served through the given communication channel as well as syntactic and structural request parsing.

The process of **selecting a web application**, as a rule, is based on the domain name and port, so, to model this process we only need to match the list of corresponding pairs (domain name, port) to each web application.

The **syntactic and structural parsing** process transforms "raw" HTTP protocol requests into data structures that are used immediately in the application logic. This parsing can be logically divided into three stages: parsing according to HTTP protocol specification, framework level parsing, and, finally, parsing according to the logic of the application itself. The first stage (parsing on the HTTP protocol level) is mostly standardized, however even on this stage, the firewall should differentiate the protocol parsing implementation nuances depending on the platform used, or otherwise a malicious attacker may use these nuances to bypass the protection system. More detailed intrusions on protocol parsing are investigated in [1].

In order to illustrate the two levels of syntax parsing, we provide an example of a body of an HTTP request for an application on the framework of Google Web Toolkit:

```
{"F":"org.gonevertical.client.re
questfactory.ApplicationRequestF
actory","I":[{"O":"wKhJ_E4nE3Sau
Zf0i01Lt8p4wEg=","P":[{"R":"1","
C":3,"T":"3cfsL4__f0SXJouPZ4iJV3
_Hzpg="}]}],"O":[{"O":"PERSIST",
"R":"1","C":3,"T":"3cfsL4__f0SXJ
ouPZ4iJV3_Hzpg=","P":{"name":"as
dasd"}}]}
```

One can see both structural data specific for GWT framework (underlined) and data specific for a particular application (**in bold**) present in the structure of the message.

In order to model the results of HTTP request parsing, a tree with noted edges and vertices is used, which is built by sequentially applying decoding steps. For instance, for the request

```
GET                      /json-
import.php?c=0&load%5B%5D=jquery
-core,jquery-
migrate&load%5B%5D=utils&ver=3.8
.2&json={%22firstName%22:%22Иван
%22,%22lastName%22:%22Иванов%22,
%22address%22:{%22postalCode%22:
101101},%22phoneNumbers%22:[%228
12123-1234%22,%22916123-
4567%22]}
```

the following parsing tree (see Fig.1) will be built by means of sequentially applying three steps of decoding.

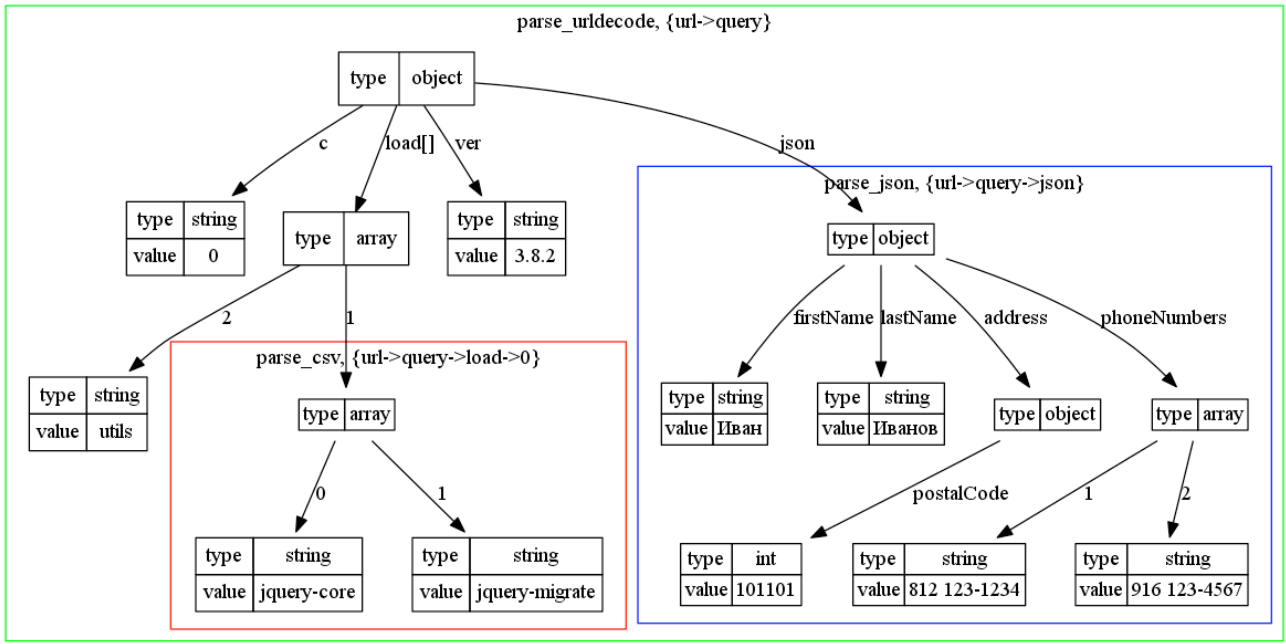

*Figure 1. Example of an HTTP request parsing tree*

In order to model the algorithm that the web application uses in parsing an arbitrary request, a solutions tree is used that describes in what order and under what conditions the decoding steps are applied (they correspond to the tops of the tree). The conditions (the tags of the arcs correspond to them) are described in the form of predicates over the contents of the operating request parsing tree. An example of the desicion tree is illustrated below (see Fig.2).

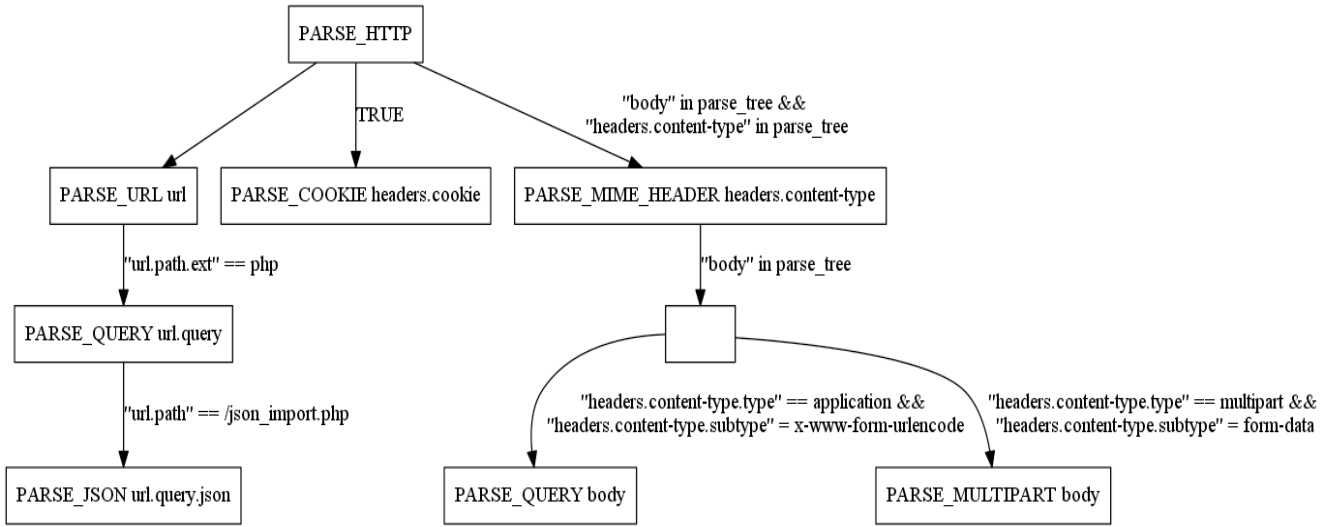

*Figure 2. Decision tree for request parsing.*

### 3.2 Actions level
On the next abstraction level, the actions level, a transition is made from the parsed HTTP requests to samples of actions. The possible quantity of all the requests to occur on an application is truly massive, however all legitame requests are merely samples of actions envisaged by the logic of the application. The quantity of different actions is, as a rule, fairly small (around 100 for the most complicated applications). The actions may have parameters that have, as a rule, a specific domain (definition area, type). Thus, any request obgliatorily sets an action and a list of its (actual) parameters. The algorithm for selecting the action and highlighting its actual parameters based on the contents of the request is sometimes referred to as routing. In order to efficiently protect a web application, a firewall must also complete a transition from requests to actions and parameters, which requires a routing modeling of the process. To ensure the syntaxical validation of the actions' parameters, the WAF must also support the syntaxical models of the parameters.

Based on an analysis of the main methods of organizing the routing process in the modern frameworks, decision lists are proposed for **routing** modelling, where the logical expressions are displayed as predicates over the contents of the parsing tree and the result is the action used and the addresses of its actual parameters in the parsing tree. We will examine the routing configuration of a Django based web application as an example:

```
urlpatterns = [
    url(r'^articles/([0-9]{4})/$',
'news.views.year_archive'),
    url(r'^articles/([0-9]{4})/([0-9]{2})/$',
'news.views.month_archive'),
    url(r'^articles/([0-9]{4})/([0-9]{2})/([0-9]+)/$',
'news.views.article_detail'),
]
```

The following decision list correspond to the example above:

| Predicate | Name of Action | Parameters |
| --- | --- | --- |
| url.path.1 = articles AND url.path.2 ~= [0-9]{4} AND url.path.3 ~= [0-9]{2} and url.path.4 ~= [0-9]+ | View News | YEAR=url.path.2, MONTH=url.path.3, ARTICLE_ID=url.path.4 |
| url.path.1 = articles AND url.path.2 ~= [0-9]{4} AND url.path.3 ~= [0-9]{2} | Past Month Archive | YEAR=url.path.2, MONTH=url.path.3 |
| url.path.1 = articles AND url.path.2 ~= [0-9]{4} | Past Year Archive | YEAR=url.path.2 |

For **syntax** modeling, we propose to differentiate between the following cases (the approach follows the scheme presented in [8]):
1. The parameter type is enumeration. In this case, the model is the list of possible parameter values.
2. The parameter corresponds to one of the a priori widespread data types (numbers, different types of indicators, e-mail addresses, phone numbers, etc.). The model in this case entails a description of the selected type (for example, in the form of a regular expression).
3. The parameter does not correspond to the predefined data type, but has a regular syntax structure. In this case, we propose to use a statistical classifier based on multi-dimentional normal distribution trained with historical data.
4. The parameter is an arbitrary formatted string entered by the user. In this case, the definition area of the parameter is unlimited and thus there is no way of creating meaningful model.

Thus, having a series of values for a parameter over a training period, one can use a set of statistical tests to determine which of the four cases listed above is abblicable, and train the corresponding classifier.

On the actions level the authentication process (including multi-level) is modeled as well as the **user session** implementation. We propose to describe these processes in terms of the life cycle of the *session attributes*. For each of these attributes we define:
- the procedure for establishing the values of an attribute (for example, the attribute "session id" may be set in the Set-Cookie header in an HTTP response, a "CSRF token" attribute – in the HTTP response body, and a "user name" attribute – in the successful authentication HTTP request body);
- the procedure for verifying the correctness of the values of an attribute transmitted in requests. For example, "session id" is transmitted in the cookie header in an HTTP request and must have a correct value for all actions in the non-public part of the service, while "CSRF token" is transmitted in the csrftoken parameter and must have a correct value for all non-idempotent (changing the condition of the application) actions;
- the procedure for invalidating the values of an attribute. For example, after executing the "exit" action, all of the values of all the attributes operating ealier are invalidated.

### 3.3 Use cases level

At the **use cases** level we analyse the logically dependend action chains – use cases. In order to model the use cases, a firewall must have the ability to describe dependencies between actions, both data dependencies (action B must operate based on the results of action A) and time dependencies (action B may be exucted only after action A is executed first).

In order to model the time dependencies, we propose to use finite state automata. Data dependency models for one of today's online banking systems are illustrated below (see Fig.3).

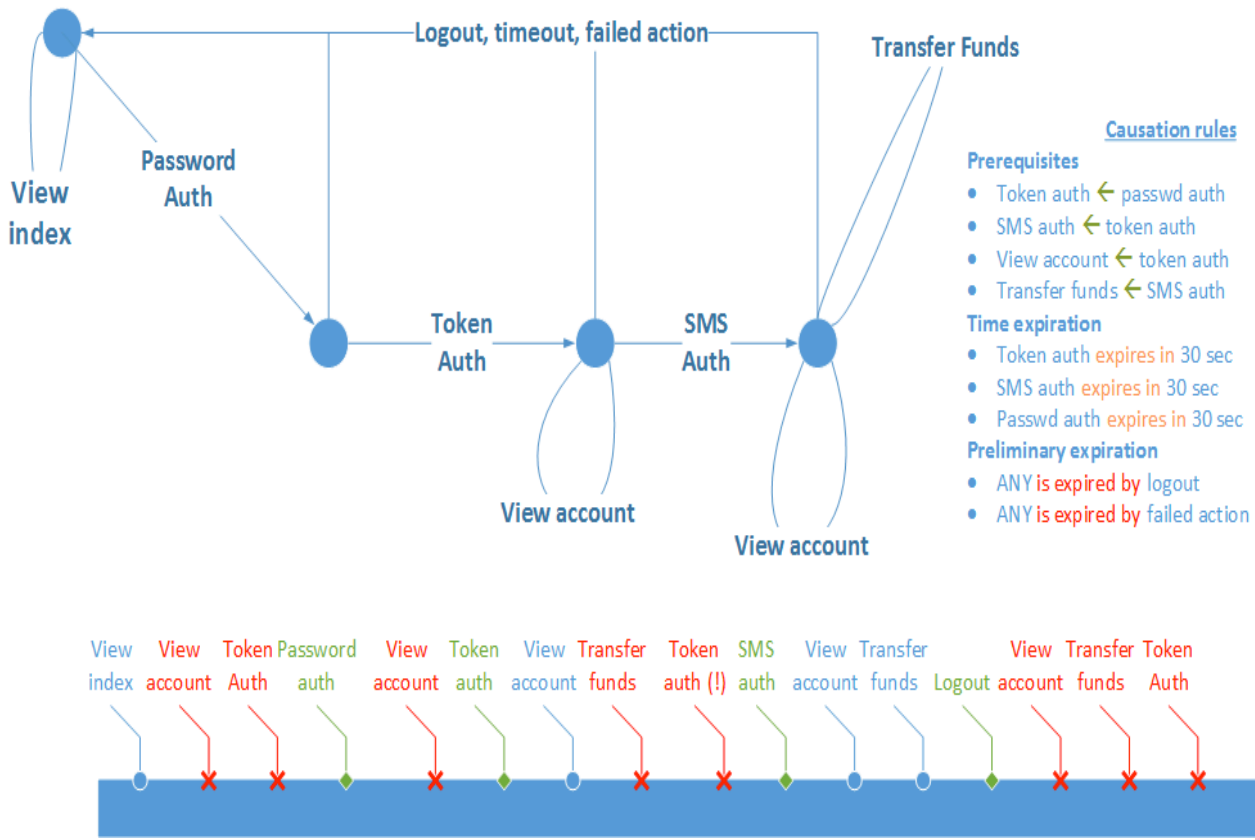

*Figure 3. Modeling dependencies between actions*

This model describes a system where initially access is only allowed to the start page (View index), then after entering user credentials (Password auth) and completing key authorization (Token auth) the user gets the opportunity to view data for the account (View account). To complete a transfer (Transfer funds), the user will also be required to enter an additional text code.

Also on the use cases level we perform a **data objects reconstruction** for learning data objects web application operates with (in the terms of its logic – for example: accounts, messages, user profiles, etc.). Additionally, the data object life cycle whould be defined: which actions correspond to the creation, viewing, modification, and removal of a certain type of object.

On the same level, the modeling process of **access control** takes place as well, the dynamic definition of a multitude of samples of actions permitted for users at each moment in time. In order to do so, we propose to analyze the contents of pages that the web server gives to one user or other and derive from them a list of actions that may be activated from the pages (envisaged by the user's interface), like in [9].

**Conclusion**

In this paper we analysed the aspects of web applications functioning from the perspective of security enforcement with the web application firewall. We proposed modeling principles of indicated aspects making it possible to create a functional model of an application and identify abnormal requests in traffic that run counter to the constructed model. The approaches used for modeling the properties of web applications allow automation with training on network data, however a description of the automation itself escapes beyond the bounds of this article. Both manual and automatic reconstruction of web application aspect model would allow WAF to adapt to the specific protected web application and dramatically improve it's accuracy.

Additional research requires modeling interaction between different web applications within the bounds of complex ecosystems (for example, "single entry point" technology", Sigle Sign-On). Also, a promising trend is the creation of user behavior models of applications to identify incidents of theft or illegal use of user credentials.

Functional modeling of web applications and taking all of its relevant properties into consideration is a necessity in creating efficient firewalls for today's web applications.